\documentclass[preprint]{aastex}
\usepackage{inputenc}
\usepackage[T2A]{fontenc}
\usepackage[english]{babel}
\usepackage{amsmath}
\usepackage{amssymb}
\usepackage{natbib}
\usepackage{bm}
\usepackage{color}

\newcommand{\vecb}[1]{{\bm{\mathrm{#1}}}}
\newcommand{\Dfrac}[2]{\frac{d#1}{d#2}}

\newcommand{\DPfrac}[2]{\frac{\partial#1}{\partial#2}}

\begin{document}
%
\title{KINEMATIC EFFECTS OF THE VELOCITY FLUCTUATIONS FOR DARK HALOS
  POPULATION IN $\Lambda$CDM MODEL}
\author{E.~P.~Kurbatov}
\affil{Institute of astronomy, Russian Academy of Sciences}
\affil{48 Pyatnitskaya st., Moscow, Russian Federation, 119017}
\email{kurbatov@inasan.ru}
\begin{abstract}
$\Lambda$CDM-model predicts an excess of dark halos
compared to the observations. The excess is seen from 
the estimates of the virialized mass inside the Local
Supercluster and its surroundings. It is shown that account for 
cosmological velocity fluctuations in the process of formation of the population of dark halos  it is possible to eliminate this
contradiction remaining within the framework of the $\Lambda$CDM cosmology. 
Based on the formalism of
Press and Schechter, we suggest a model of formation of dark halo population 
which takes into account 
the kinematic effects in the dark matter. The model allows  
a quantitative explanation for the observable deficit of the virialized mass in the local Universe.
\end{abstract}

\section{Introduction}

The inconsistencies between the predictions of the $\Lambda$CDM cosmological model 
and the
observations concern the distribution of the matter on the scales smaller than the
scale of homogeneity of the Universe. In particular, note the deficit of
virialized matter in the local
Universe~\citep{Makarov2011MNRAS.412.2498M}. Counts of the mass of the galaxies,
virialized groups, and clusters of galaxies in the Local Supercluster and its environs
performed by many authors revealed the lack of mass inside the virialized
objects, amounting to a half of an order of magnitude (see the references
in~\citet{Makarov2011MNRAS.412.2498M}). According to the \citet{Makarov2011MNRAS.412.2498M}
catalog the estimate of the local density
parameter inside a sphere of the radius $48$~Mpc is about $0.08 \pm 0.02$. As a
possible explanation of this deficit \citet{Makarov2011MNRAS.412.2498M}
suggested that about $2/3$ of the matter is located 
outside the virialized areas. Instead, this dark
matter is either concentrated in dark clumps, outside the virial regions, or distributed diffusively. In the
favor of the idea of the dark clumps tell observational data on weak
lensing~\citep{Jee2005ApJ...618...46J} and on disturbed
galaxies~\citep{Karachentsev2006A&A...451..817K}.

One of the generally accepted ideas on the Large-Scale Structure formation in
the Universe is the hierarchical model. Formation of the virialized halos
population in the $\Lambda$CDM model is represented as a continuous process of
condensing and clustering of the structures which develop from the density
perturbations. Stochastic nature of this process is determined by the
properties of the initial cosmological fluctuations. As a result, a
hierarchical halos structure forms, which consists of galactic clusters,
virialized groups, field galaxies and low-mass satellite galaxies. \citet{Press1974ApJ...187..425P} suggested a simple model for
evolution of the dark halos mass function. Development of this model 
in later studies  helped to solve the problem to a good approximation. The
extension of the PS model known as the Excursion-Set or the Extended PS (EPS)
formalism \citep{Bond1991ApJ...379..440B,Lacey1993MNRAS.262..627L} is based
on two
assumptions: (i) the requirement for the perturbation to be virialized at a
given time which can be formulated in the terms of the density perturbation field at
the linear stage of its evolution; (ii) the mass function of halos is formulated for
those objects which are at the top of the hierarchical structure, i.e. they
do not belong to other halos. In this formulation, the problem can be solved
using the linear perturbation theory only.

In the formation process of the dark halos population the environment effects,
which can lead to the loss of the halo mass or to the
ejection of the sub-halos from the parent clusters,
have an essential importance. In the numerical model of
\citet{Diemand2007ApJ...667..859D} was shown that the most intensive  loss
of sub-halo
happens before virialization of the parental halo. An already formed halo can
accumulate  up to $20\%$ of its mass by the accretion of surrounding dark matter.

Despite of the quite obvious role of the environment effects, they are not the
only ones which are able to influence the halo population. For instance, 
\citet{Valageas2012PhRvD..86l3501V} have shown that the velocity fluctuations in the
dark matter (the author considered the Warm Dark Matter model, WDM) can affect
the density fluctuations statistics on scales
$\lesssim 0.1 h^{-1}$~Mpc. \citet{Valageas2012PhRvD..86l3501V} did not  find any considerable effect of the
velocity fluctuations for the dark halos population. It should be emphasized however that the calculations were constrained to corrections to the power spectrum only.
More, approach used in the latter study  required  modification of the
$\Lambda$CDM-model.

An interesting problem is the possibility of direct kinematic effect of the
velocity fluctuations on the process of the formation of the individual
halos as well as their population. It is well known that the velocity fluctuations
grow along to the density fluctuations~\citep{Crocce2006PhRvD..73f3519C}. Due
to the moment conservation, the halo during formation process inherits
velocity of the dark matter averaged over its scale. When
this halo becomes involved in the formation process of the larger scale
gravitational condensation, its velocity may be large enough to leave the parental
structure. This effect may change the history of evolution of dark halos,
the  dark halos population itself,
as well as the history of the chemical evolution of the galaxies. 

In this paper, a method for accounting for the kinematic effects in the dark halos population 
process is suggested. The evolutionary model of the population is
formulated on the base of the Excursion-Set
theory~\citep{Bond1991ApJ...379..440B,Lacey1993MNRAS.262..627L} in terms of
kinetic equation. Also, the  problems of the choice of the initial conditions and
of the influence of the background structure are considered.

In \S~2 the EPS model is briefly examined. In \S~3 formation
model for individual halo with kinematic effects is suggested and the
kinetic equation for the mass function is obtained. In \S~4 the model is
applied to the problem of the virialized mass deficit in the local
Universe. Conclusions are presented in \S~5.

\section{Formation of the halo population in the EPS model}

\subsection{Growth of the cosmological perturbations}

In the $\Lambda$CDM model the Large-Scale Structure of the Universe is formed from
the growing fluctuations of density and velocity in the cold dark
matter. Evolution law of the structures with characteristic mass $m$ can be
linked to the properties of the overdensity field
$\delta \equiv (\rho - \rho_\mathrm{cr,0})/\rho_\mathrm{cr,0}$ averaged over the
volume containing mass $m$. The averaging is defined as a convolution of the field
with a filter $W$:
\begin{equation}
  \label{eq:filtered_overdensity}
  \delta(a, \vecb{x}, m)
  = \int d^3x'\,W(\vecb{x}' - \vecb{x}, m)\,\delta(a, \vecb{x}')  \;,
\end{equation}
where $a$ is the scale factor, which is related to the redshift $z$ as
$a = 1/(z+1)$; $m$ is the halo mass. The same in the terms of the Fourier
transform is%
\footnote{%
The Fourier transform is defined as
$\tilde f(\vecb{k}) = (2\pi)^{-3/2} \int d^3x\,e^{i\vecb{kx}} f(\vecb{x})$.
}:
\begin{equation}
  \tilde{\delta}(a, \vecb{k}, m)
  = (2\pi)^{3/2}\,\tilde{W}(\vecb{k}, m)\,\tilde{\delta}(a, \vecb{k})  \;.
\end{equation}
The \glqq{top-hat}\grqq\ filter will be adopted hereinafter:
\begin{equation}
  W(\vecb{x}, m) = \frac{3}{4\pi X^3}\,\theta(X - x)  \;,
\end{equation}
\begin{equation}
  (2\pi)^{3/2}\,\tilde{W}(\vecb{k}, m) =
  3\,\frac{\sin kX - kX \cos kX}{k^3 X^3}  \;,
\end{equation}
where $X = [ 3 m / (4\pi \Omega_\mathrm{m,0} \rho_\mathrm{cr,0}) ]^{1/3}$.

The random fluctuations field is assumed to be Gaussian with zero mean and
delta-correlated over Fourier modes,
\begin{equation}
  \langle \tilde\delta(a, \vecb{k})\,\tilde\delta^\ast(a, \vecb{k}') \rangle =
  \delta_\mathrm{D}(\vecb{k} - \vecb{k}')\,P(a, k)\;,
\end{equation}
where $\delta_\mathrm{D}$ is the Dirac's delta-function, $P(a, k)$ is the
overdensity power spectrum at the moment corresponding to the scale factor $a$.
Thereby, over  spatial scale associated with the given mass $m$ the overdensity
field can be characterized by only the variance value, which is independent of
spatial position:
\begin{equation}
  \label{eq:overdensity_variance}
  S(a, m) \equiv \langle \delta^2(a, \vecb{x}, m) \rangle
  = \int d^3k\,|\tilde{W}(\vecb{k}, m)|^2\,P(a, k)  \;.
\end{equation}

Velocity divergence fluctuations field
$\theta \equiv \nabla_\vecb{x} (a \dot{\vecb{x}})$
has the same properties:
\begin{equation}
  \langle \tilde\theta(a, \vecb{k})\,\tilde\theta^\ast(a, \vecb{k}') \rangle
  = \delta_\mathrm{D}(\vecb{k} - \vecb{k}')\,P_\theta(a, k)  \;,
\end{equation}
where $P_\theta(a, k)$ is the power spectrum of the field
$\theta(a, \vecb{x})$. Let denote as $\vecb{v} \equiv a \dot{\vecb{x}}$ the 
field of the
physical velocity determined relatively to the expanding Universe. We neglect the
vortical part of the velocity fluctuations. In this case, the velocity field is
determined by the divergence field only:
\begin{equation}
  \langle \tilde\theta(a, \vecb{k})\,\tilde\theta^\ast(a, \vecb{k}') \rangle =
  \sum_{j,j'} k_j k'_{j'}\,
  \langle \tilde{v}_j(a, \vecb{k})\,\tilde{v}_{j'}^\ast(a, \vecb{k}') \rangle =
  \delta_\mathrm{D}(\vecb{k} - \vecb{k}')
  \sum_j k_j^2\,\langle |\tilde{v}_j(a, \vecb{k})|^2 \rangle  \;.
\end{equation}
Let define as $P_v(a, k) \equiv \langle |\tilde{v}_j(a, \vecb{k})|^2 \rangle$ 
the power
spectrum of the velocity in the given spatial direction (the velocity
distribution is assumed to be isotropic). Then the power spectrum of the field
$\vecb{v}$ will be
\begin{equation}
  P_v(a, k) = \frac{P_\theta(a, k)}{k^2}  \;,
\end{equation}
where $k^2 = |\vecb{k}|^2$.

At large redshifts the overdensity and velocity divergence amplitudes evolve
according to the linear law \citep{Peebles1980lssu.book.....P,Crocce2006PhRvD..73f3519C}:
\begin{align}
  \label{eq:linear_overdensity}
  \tilde{\delta}(a, \vecb{k}) &=
  \frac{D}{D_\mathrm{i}}\,\tilde{\delta}(a_\mathrm{i}, \vecb{k}) \equiv
  D \tilde{\delta}_\mathrm{L}(\vecb{k})  \;,  \\
  \label{eq:linear_veldiv}
  \tilde{\theta}(a, \vecb{k}) &=
  \frac{a H f D}{a_\mathrm{i} H_\mathrm{i} f_\mathrm{i} D_\mathrm{i}}\,
  \tilde{\theta}(a_\mathrm{i}, \vecb{k}) \equiv
  a H f D \tilde{\theta}_\mathrm{L}(\vecb{k})  \;,
\end{align}
where $H = H(a)$ is the Hubble parameter, $D = D(a)$ is the linear growth
factor, $f \equiv d\ln D/d\ln a$, index \glqq{$\mathrm{i}$}\grqq\ marks the
initial time. Hereinafter, by the index \glqq{$\mathrm{L}$}\grqq\ we will
denote the values reduced to the unit growth factor, i.e. not depending on the
redshift. For linear approximation for the amplitudes, the variances are
subject to the quadratic growth law:
\begin{align}
  \label{eq:overdensity_variance}
  S(a, m) &= D^2 S_\mathrm{L}(m)  \\
  \label{eq:velocity_variance}
  S_v(a, m) &= D_v^2 S_{v,\mathrm{L}}(m)  \;,
\end{align}
where $D_v \equiv a H f D$. Physically, interesting is the situation when
the initial overdensity and velocity divergence perturbations are
proportional to each other. It can be shown~\citep{Crocce2006PhRvD..73f3519C}
that in this case
\begin{equation}
  \tilde{\delta}(a_\mathrm{i}, \vecb{k}) =
  - \frac{\tilde{\theta}(a_\mathrm{i}, \vecb{k})}%
  {a_\mathrm{i} H_\mathrm{i} f_\mathrm{i}}  \;,
\end{equation}
i. e., the power spectra of the fields $\tilde{\delta}_\mathrm{L}(\vecb{k})$
and $\tilde{\theta}_\mathrm{L}(\vecb{k})$ coincide.

Due to the momentum conservation, the forming halo as a whole inherits the
velocity of the proto-halo matter, namely, velocity, averaged over 
proto-halo mass scale
\begin{equation}
  \vecb{v}(a, \vecb{x}, m)
  = \int d^3x'\,W(\vecb{x}' - \vecb{x}, m)\,\vecb{v}(a, \vecb{x}')  \;.
\end{equation}
Let assume, that a proto-halo with mass $M$ had  velocity
$\vecb{V}(a, \vecb{X}, M)$. Then the variance of a particular 
spatial component of the
velocity $v_j(a, \vecb{x}, m)$ of a sub-halo with the mass $m$, to the  first
approximation will depend neither on position of the sub-halo inside the
proto-halo nor the velocity of the proto-halo,  but only on 
the variance of the velocity field on the scale $M$%
\footnote{%
Closer the centers of the volumes $M$ and $m$ are to each other, 
the better this approximation is~\citep{Hoffman1991ApJ...380L...5H}. See
also~\citep{Kurbatov2014ARep...58..386K}.}:
\begin{equation}
  \label{eq:relative_velocity_variance}
  S_v(a, m, M) \equiv
  D_v^2\,[S_{v,\mathrm{L}}(m) - S_{v,\mathrm{L}}(M)]  \;.
\end{equation}

\begin{figure}[!ht]
  \centering
  \includegraphics{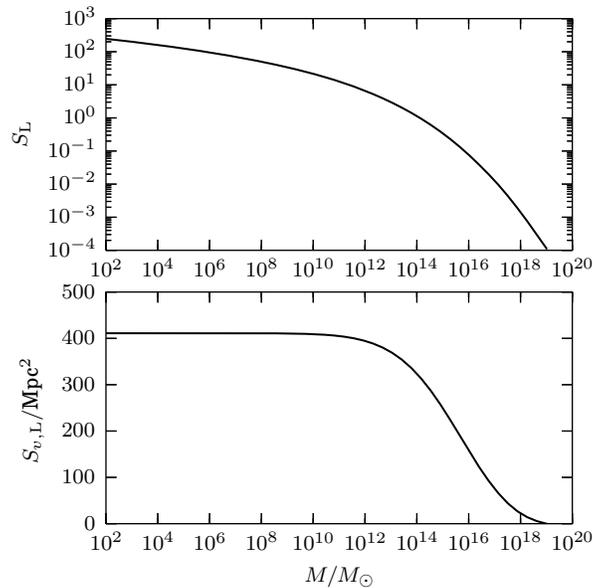}
  \caption{Dependence of the variance of overdensity fluctuations (upper panel) and the variance of velocity fluctuations (lower panel) on the mass scale.}
  \label{fig:variances}
\end{figure}

In the present paper the $\Lambda$CDM model parameters obtained by the Planck
mission~\cite{PlanckCollaboration2013arXiv1303.5076P} were adopted:
$h = 0.673$, $\Omega_\mathrm{m,0} = 0.315$,
$\Omega_{\Lambda,0} = 1 - \Omega_\mathrm{m,0} = 0.685$, $\sigma_8 = 0.828$,
$n_\mathrm{S} = 0.9603$. The power spectrum of the overdensity fluctuations was
calculated for these parameters by means of the code \textsc{CAMB}
\citep{Lewis2000ApJ...538..473L} 
 for wavenumbers from $10^{-3}$~Mpc$^{-1}$ to
$2$~Mpc$^{-1}$ (on-line interface to the \textsc{CAMB} 
is presented in the LAMBDA project \citeyear{LAMBDA2013}). 
Outside of this range, the theoretical power spectrum was used with
the power index $n_\mathrm{S}$ and the transfer function \citep[][p. 60]{Bardeen1986ApJ...304...15B} with the shape parameter
$\Gamma \equiv h^2 \Omega_\mathrm{m,0} = 0.14$~Mpc$^{-1}$. Plots of the
overdensity and velocity variances calculated in such a  model
are shown in the Fig.~\ref{fig:variances}.
The linear growth factor was adopted 
from~\citet{Bildhauer1992A&A...263...23B} using normalization
$D(a \approx 0) = a$.

\subsection{Evolution of the mass function}

The EPS model \citep{Peacock1990MNRAS.243..133P,Bond1991ApJ...379..440B} can be
formulated with the excursion set approach as follows. Consider the overdensity
fluctuations field $\delta(a, \vecb{x}, m)$ with the mass scale $m$. If the
perturbation is spherical, it can be argued that its amplitude on the linear
stage is determined by the initial amplitude
only~\citep{Peebles1980lssu.book.....P}, or, equivalently, by
$\delta_\mathrm{L}(\vecb{x}, m)$. This feature can be used to choose such
perturbations which collapse and virialize to the halo stage by a given
redshift. It's necessary to note, however, that some already virialized halos may be
absorbed by more massive ones. Therefore, the only initial perturbations accounted
in the mass function for certain redshift should be the most massive ones with
required virialization time. 

These ideas may be represented as a typical problem of theory of the
stochastic processes. A stochastic process%
\footnote{%
Hereafter the index \glqq{$\mathrm{L}$}\grqq\ will be omitted as the redshift
dependency will be denoted explicitly.
}
$\delta(S)$ starts from the point $\delta = 0$ with the parameter  $S = 0$,
which corresponds to $m = \infty$. Its evolution obeys equation
\begin{equation}
  \label{eq:random_walk}
  \Dfrac{\delta}{S} = \eta  \;,
\end{equation}
where $\eta = \eta(S)$ is a Gaussian random process with zero mean, unit variance, and the
correlation function
$\langle{\eta(S)\,\eta(S')}\rangle = \delta_\mathrm{D}(S - S')$
\citep{Maggiore2010ApJ...711..907M}. The mass $m$ corresponding to the value of the
parameter $S$, when a certain threshold $\delta_\mathrm{c}/D(z)$ is reached by the
process $\delta(S)$ for the first \glqq{time}\grqq\ $S$, is interpreted as the mass of the halo virialized by
the given redshift $z$. The  distribution function of these masses is the halo
mass function sought for.

The value $\delta_\mathrm{c}$ is the linear amplitude
of the spherical overdensity perturbation (\ref{eq:linear_overdensity})
collapsing at the moment $z$. In the spherical collapse
model~\citep{Peebles1980lssu.book.....P} this value reduced to the unit growth
factor is approximately $1.686$. In our case%
\footnote{%
In models of the EPS class the linear growth factor $D$ is usually normalized
to take unit value at $z = 0$~\citep{Lacey1993MNRAS.262..627L}. In the present
paper the alternative normalization is used: $D(z) \approx 1/(z + 1)$ at
$z \gg 1$. For cosmological parameters adopted here (see previous Sec.) we have
$D(z = 0) \approx 0.788$.
}
\begin{equation}
  \label{eq:threshold_overdensity}
  \delta_\mathrm{c} = 1.686\,D(z = 0)  \;.
\end{equation}
The resulting halo mass function can be represented as a probability
density~\citep{Bond1991ApJ...379..440B,Lacey1993MNRAS.262..627L}
\begin{equation}
  \label{eq:eps_mass_function}
  f^{(0)}(z, S) = \frac{\omega}{\sqrt{2\pi S^3}}\,
  \exp\left( -\frac{\omega^2}{2 S} \right)
\end{equation}
or the cumulative distribution function
\begin{equation}
  \label{eq:eps_mass_function_cumulative}
  F^{(0)}(z, < S) = \int_0^S dS'\,f^{(0)}(z, S') =
  \operatorname{erfc}\left( \frac{\omega}{\sqrt{2 S}} \right)  \;,
\end{equation}
where notation $\omega \equiv \delta_\mathrm{c}/D(z)$ is introduced.

The variance $S$ or the mass of a halo with a largest formation
rate at the given redshift correspond to the maximum of the function
$\partial f^{(0)}/\partial\omega$ and thus obeys equation
$\omega^2 = (3 + \sqrt{6})\,S$. Most  of the halos formed per unit time has
the variance in the interval $\log S = \log[\omega^2/(3 + \sqrt{6})] \pm 0.5$. At
large redshifts, the interval can be estimated as
\begin{equation}
  \label{eq:variance_interval}
  S \approx (0.15 \dots 1.5)\,(z + 1)^2  \;.
\end{equation}

It should be noted that the process $\delta(S)$ defined by 
Eq.~(\ref{eq:filtered_overdensity}) for an arbitrary filter strictly speaking does not comply
to Eq.~(\ref{eq:random_walk}) where the r.h.s. is delta-correlated. That is, the
process is not Markovian. Moreover, the spherical collapse model is quite a 
rough approximation, neither any possible deviations from the spherical symmetry
 in the initial conditions nor tidal action of
environment are  considered. These issues can be resolved by corrections to the excursion set
formalism~\citep{Bond1991ApJ...379..440B,Maggiore2010ApJ...711..907M,%
Maggiore2010ApJ...717..515M}. However in the present paper we  base on the
simple case of the excursion set theory.

\section{Kinematic effects in the dark matter}

\subsection{Formation of a single halo}

Transition of a spherical perturbation into a halo proceeds  in four stages:
initial expansion, detachment from the Hubble flow, compression,
virialization~\citep{Peebles1980lssu.book.....P}. The moment of detachment or
turnaround point is located approximately in the middle between the start of the
expansion and the virialization, when the formation is
completed~\citep{Peebles1980lssu.book.....P}. In the spherical collapse model,
the halo formation is described by motion of disjoint layers or test particles
in the gravitational field of a point mass. At that, the mass enclosed inside the layer
is conserved. For the perturbation to become collapsed at a given redshift
$z_\mathrm{f}$, its linear overdensity must grow
as~\citep{Lacey1993MNRAS.262..627L}
\begin{equation}
  \Delta(z) =
  \frac{D(z)}{D(z_\mathrm{f})}\,\frac{\delta_\mathrm{c}}{D(z = 0)}  \;.
\end{equation}

It was mentioned above that sub-halos have random velocities relatively to the
parent halo. Let consider the possibility that the velocity of the sub-halo is large
enough to leave the proto-halo. Let's write an expression for mechanical
energy of a test particle inside the parent proto-halo using physical
coordinates, and neglecting the $\Lambda$-term:
\begin{equation}
  E = \frac{\dot{\vecb{r}}^2}{2} + \Phi  \;.
\end{equation}
At large redshifts, the overdensity amplitude of the parent proto-halo is
low, thus the density distribution inside the parent proto-halo may be considered as uniform.
Gravitational potential in this case is
\begin{equation}
  \Phi = - \frac{G M}{R}
  \left( \frac{3}{2} - \frac{r^2}{2 R^2} \right)  \;,
\end{equation}
where $M$ is the proto-halo mass; the scale $R$ is defined as a physical radius of
the proto-halo at the turnaround, containing the mass $M$:
\begin{equation}
  M = \frac{4\pi}{3}\,\rho_\mathrm{cr,0} \Omega_\mathrm{m,0}\,
  (1 + \Delta)\,\frac{R^3}{a^3}  \;.
\end{equation}
We formulate escape condition for the particle as the constraints for the
energy value, $E > 0$:
\begin{equation}
  \label{eq:escape_condition}
  (a \dot{x} + a H x)^2 >
  \frac{H_0^2 \Omega_\mathrm{m,0}}{a} X^2\,(1 + \Delta)
  \left( \frac{3}{2} - \frac{x^2}{2 X^2} \right)  \;,
\end{equation}
where $X = R/a$, $x = r/a$, $x \leqslant X$, $a < a_\mathrm{f}$. Condition
(\ref{eq:escape_condition}) defines the lower bound of the velocity sufficient
 to leave the proto-halo:
\begin{equation}
  \label{eq:escape_condition_explicit}
  \dot{x} > - H x
  + \left[ \frac{H_0^2 \Omega_\mathrm{m,0}}{a^3}\,X^2\,(1 + \Delta)
    \left( \frac{3}{2} - \frac{x^2}{2 X^2} \right)  \right]^{1/2}  \;.
\end{equation}

Let's estimate the escape probability. Assume the following: (i) the test
particle's location is near the edge of the proto-halo; (ii) the proto-halo
collapses at a high redshift
($H \approx H_0 \Omega_\mathrm{m,0}^{1/2} a^{-3/2}$); (iii) the amplitude of
the proto-halo is low ($\Delta \ll 1$). Then the escape condition
(\ref{eq:escape_condition_explicit}) takes the form
\begin{equation}
  \dot{x} > \frac{1}{2}\,H X \Delta  \;.
\end{equation}
Sub-halos served as the test particles in our formulation of the problem.
Let's multiply the last inequality by the scale factor $a$, then
substitute into the l.h.s. the most probable value of the relative velocity
modulus of the sub-halo inside the proto-halo, assuming the Gaussian
distribution with zero mean and the variance
(\ref{eq:relative_velocity_variance}). Also, let express $X$ via the proto-halo
mass. In the high-$z$ limit, the condition (\ref{eq:escape_condition_explicit})
takes the form
\begin{equation}
  \label{eq:escape_cond_rough}
  [ S_v(m) - S_v(M) ]^{1/2} \gtrsim
  3\,(z_\mathrm{f} + 1) \left( \frac{M}{10^{14} M_\odot} \right)^{1/3}
  \;\text{Mpc}  \;.
\end{equation}
It can be shown (see Appendix) that in not very strict conditions ($m \to 0$,
$M < 10^{14} M_\odot$) the l.h.s. of the last inequality can be estimated as
\begin{equation}
  [ S_v(0) - S_v(M) ]^{1/2} \approx
  1.75\,\{ \dots \} \left( \frac{M}{10^{14} M_\odot} \right)^{1/3}
  \;\text{Mpc}  \;,
\end{equation}
where expression in braces is slowly changing decreasing function of $M$
(it becomes close to  $200$ when $M = 10^2 M_\odot$, and $1$ when
$M = 10^{14} M_\odot$). Adopting (\ref{eq:escape_cond_rough}), we get
(\ref{eq:escape_condition_explicit}) in the form
\begin{equation}
  \{ \dots \} \gtrsim z_\mathrm{f} + 1  \;.
\end{equation}
This inequality formally shows that for any $z_\mathrm{f}$ there are
perturbations which masses are small enough to obey the escape
condition for its sub-halos. It is necessary to remember, however, that for any redshift there is a certain
mass interval in which most of the halos form, see
Eq.~(\ref{eq:variance_interval}). For sub-halos escape to have an effect upon the
population formation process, the proto-halo mass must satisfy both Eqs.
(\ref{eq:variance_interval}) and (\ref{eq:escape_cond_rough}). This requirement
can be represented as
\begin{equation}
  \frac{S_v(0) - S_v(M)}{(M/10^{14} M_\odot)^{2/3}\;\text{Mpc}^2\:} \gtrsim
  S(M)  \;.
\end{equation}
Let's estimate both sides of this inequality in the low-mass limit. For
simplicity let use the \glqq{k-sharp}\grqq\ filter for which the Fourier image is
$\tilde{W}(\vecb{k}, M) = \theta(1 - k X)$. Then the overdensity and the
velocity variances are
\begin{equation}
  S(M) \propto \int_0^{X^{-1}(M)} dk\,k^2 P(k)
\end{equation}
and
\begin{equation}
  S_v(0) - S_v(M) \propto \int_{X^{-1}(M)}^\infty dk\,P(k)  \;.
\end{equation}
Calculating the differentials, it's easy to show that
\begin{equation}
  \Dfrac{}{S} \frac{S_v(0) - S_v(M)}{M^{2/3}} \propto
  -1 + \frac{2X }{P(X^{-1})} \int_{X^{-1}}^\infty dk\,P(k)  \;,
\end{equation}
where the proportionality factor is a positive constant. We assumed also
$k \equiv X^{-1}$. Defining the power spectrum as $P(k) \equiv \Pi(k)/k^3$, 
by formal integration of the r.h.s we obtain
\begin{equation}
  \label{eq:variances_differentiation}
  \Dfrac{}{S} \frac{S_v(0) - S_v(M)}{M^{2/3}} \propto
  \frac{1}{X^2 \Pi(X^{-1})} \int d\Pi'\,X'^2 \geqslant 0  \;.
\end{equation}
If $\Pi$ is a constant (this is true in a rough approximation) then the r.h.s. of
(\ref{eq:variances_differentiation}) is zero. In a more strict approximation we
have $\Pi \propto \ln^2(k/\Gamma)$ i.e. slowly growing function of the
wavenumber~\citep{Bardeen1986ApJ...304...15B}. The conclusion is: the lesser
proto-halo mass is, the lesser the probability for its sub-halos to escape. Given
that the formation of the structures with time goes from the low masses to
the larger ones, it can be argued that in the high redshift limit the sub-halo
escape does not affect the formation of the dark halo population.

Let's denote as $\sigma_v \equiv [ S_v(m) - S_v(M) ]^{1/2}$ the standard deviation
of the 1-D velocity (reduced to the unit growth factor) of a sub-halo $m$
relatively to a proto-halo $M$. The probability for modulus of the Gaussian distributed sub-halo velocity to exceed some $v$ is
\begin{equation}
  P_v(> v) = \operatorname{erfc}\left( \frac{v}{\sqrt{2}\,\sigma_v} \right)
  + \frac{2}{\sqrt{\pi}}\,\frac{v}{\sqrt{2}\,\sigma_v}\,
  \exp\left( -\frac{v^2}{2 \sigma_v^2} \right)  \;.
\end{equation}
Given the uniform spatial distribution of the sub-halos at the early stage of
the proto-halo evolution, the probability for sub-halo $m$ to escape 
proto-halo $M$ formed at
the redshift $z_\mathrm{f}$ is
\begin{equation}
  \label{eq:escape_fraction}
  \chi_\mathrm{ej}(z_\mathrm{f}, M, m) \equiv
  \frac{3}{X^3} \int_0^X dx\,x^2 P_v(> v_\mathrm{ej})  \;,
\end{equation}
where a threshold velocity value $v_\mathrm{ej}$ is the r.h.s. of
(\ref{eq:escape_condition_explicit}) reduced to the unit growth factor:
\begin{equation}
  v_\mathrm{ej} =
  - \frac{a H}{D_{v}}\,x
  + \left[
    \frac{H_0^2 \Omega_\mathrm{m,0}}{a D_{v}^2}\,X^2\,
    (1 + \Delta) \left( \frac{3}{2} - \frac{x^2}{2 X^2} \right)
    \right]^{1/2}  \;.
\end{equation}

\begin{figure}[!ht]
  \centering
  \includegraphics{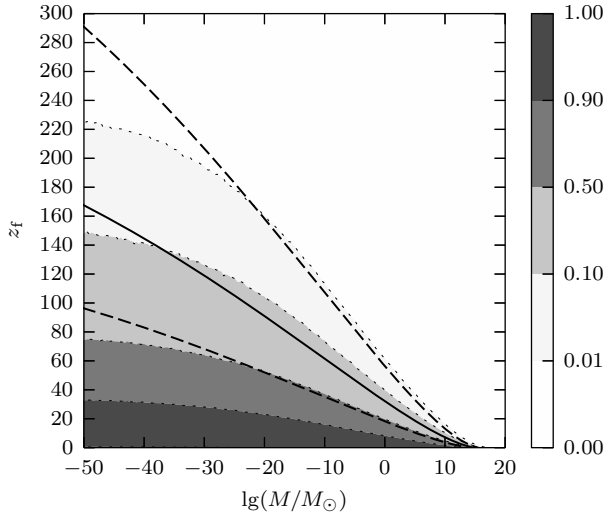}
  \caption{Isolines of the probability of ejection of sub-halos as a function of proto-halo mass (horizontal axis) and redshift (vertical axis). The probability is estimated using Eq.~(\ref{eq:escape_cond_rough}) for sub-halo masses $m \to 0$. The solid curve marks the era of the most intensive formation proto-halo of a given mass. The primary mass interval of the proto-halos formed at a given epoch is bounded by the dashed lines.}
  \label{fig:ejfrac_lowm}
\end{figure}

\begin{figure}[!ht]
  \centering
  \includegraphics{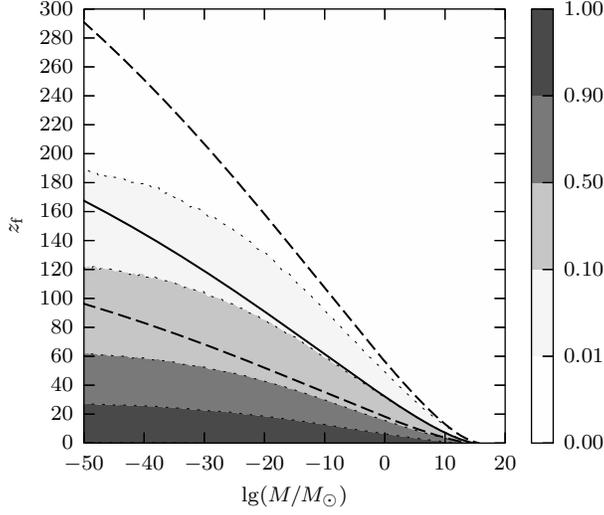}
  \caption{Same as in the Fig.~\ref{fig:ejfrac_lowm}, but the sub-halo masses are one-third of the proto-halo mass.}
  \label{fig:ejfrac_highm}
\end{figure}

Distribution of $\chi_\mathrm{ej}$ for the sub-halos of low masses is shown in
Fig.~\ref{fig:ejfrac_lowm}.
Comparing the position of contour lines and the mass interval of proto-halos
formed by a given era (the mass interval is bounded by thick dashed lines), one can 
see that for the formation redshifts $z_\mathrm{f} \lesssim 100$ the mass
fraction of escaping sub-halos of low masses is approximately constant for
all $z_\mathrm{f}$ and can be estimated as $30\%$, roughly. The same is true in
the case when the sub-halo mass is one-third of that of the proto-halo, but the
escaping mass fraction is about $15\%$ (see Fig.~\ref{fig:ejfrac_highm}).
Thereby, a significant fraction of the sub-halos can escape the parent
proto-halo, thus avoiding the absorption.

\subsection{Evolution of the mass function with account for  kinematic effects}

In the standard excursion set approach a mass inside each collapsed
perturbation is conserved during the collapse and equals to the subsequent halo
mass. This suggestion makes realizations of the random process $\delta(S)$
independent up to some small corrections noted above. As it became clear,
proto-halos may loss a significant fraction of the sub-halos resulting in
transition of the sub-halos to proto-halo of larger scales which undergo
collapse at later times. Consequently, the values of $\delta(S)$ will correlate
both in scales and redshifts. This fact makes it difficult to use the excursion
set formalism explicitly.

Greater clarity can have an approach based on the kinetic equation, when the
evolution of the mass function  explicitly presented as a result of the
absorption process of the sub-halo by the large-scale perturbations. Let's
write the total probability of a halo formation at the moment $\omega_1$, the mass of which  corresponds to the interval $(S_1, S_1 + dS_1)$, and another halo
$(S_2, S_2 + dS_2)$ at the moment $\omega_2$, given conditions $S_1 > S_2$ and
$\omega_1 > \omega_2$:
\begin{align}
  \label{eq:joint_prb_sink}
  p^{(0)}(\omega_1, S_1;\,\omega_2, S_2)\,dS_1 dS_2 &={}
  f^{(0)}(\omega_1, S_1\,|\,\omega_2, S_2)\,
  f^{(0)}(\omega_2, S_2)\,dS_1 dS_2 =  \\
  \label{eq:joint_prb_source}
  &{}= f^{(0)}(\omega_2, S_2\,|\,\omega_1, S_1)\,
  f^{(0)}(\omega_1, S_1)\,dS_1 dS_2  \;,
\end{align}
where $f^{(0)}(\omega, S)$ is the PS mass function
(\ref{eq:eps_mass_function}). Reducing the  total
probability, one obtains
\begin{equation}
  \int_{S_2}^\infty dS_1\,p^{(0)}(\omega_1, S_1;\,\omega_2, S_2) =
  f^{(0)}(\omega_2, S_2)  \;,
\end{equation}
\begin{equation}
  \int_0^{S_1} dS_2\,p^{(0)}(\omega_1, S_1;\,\omega_2, S_2) =
  f^{(0)}(\omega_1, S_1)  \;.
\end{equation}
These relations suggest  a possible form of the kinetic equation:
\begin{multline}
  f^{(0)}(\omega, S) - f^{(0)}(\omega + \Delta\omega, S) ={}  \\
  {}= \int_S^\infty dS'\,p^{(0)}(\omega + \Delta\omega, S';\,\omega, S)
    - \int_0^S dS'\,p^{(0)}(\omega + \Delta\omega, S;\,\omega, S')  \;,
\end{multline}
i.e.
\begin{multline}
  \label{eq:kinetic_ps_exact}
  - \DPfrac{f^{(0)}(\omega, S)}{\omega} ={}  \\
  {}= \lim_{\Delta\omega \to 0+} \frac{1}{\Delta\omega}
  \left[
    \int_S^\infty dS'\,p^{(0)}(\omega + \Delta\omega, S';\,\omega, S)
    - \int_0^S dS'\,p^{(0)}(\omega + \Delta\omega, S;\,\omega, S')
  \right]  \;.
\end{multline}
There is a \glqq{minus}\grqq\ sign in the l.h.s. of 
Eq.~(\ref{eq:kinetic_ps_exact}) because the $\omega$ parameter decreases with
the time. Terms in the r.h.s. are the source and the sink respectively.

According to the total probability formula (\ref{eq:joint_prb_sink},
\ref{eq:joint_prb_source}) the source and the sink can be written in
different ways. We choose the way which leads to the most familiar form of
kinetic equation: 
\begin{align}
  \label{eq:kinetic_ps_exact_usual}
  - \DPfrac{f^{(0)}(\omega, S)}{\omega} =
  \lim_{\Delta\omega \to 0+} \frac{1}{\Delta\omega}
  \bigg[
    & \int_S^\infty dS'\,f^{(0)}(\omega, S\,|\,\omega + \Delta\omega, S')\,
    f^{(0)}(\omega + \Delta\omega, S') -{}  \\
    &{}- f^{(0)}(\omega + \Delta\omega, S)
    \int_0^S dS'\,f^{(0)}(\omega, S'\,|\,\omega + \Delta\omega, S)
    \bigg]  \;.
\end{align}
\glqq{Transfer cross section}\grqq\ here is the conditional PDF
$f^{(0)}(\omega, S\,|\,\omega + \Delta\omega, S')$, and the unknown function is
$f^{(0)}(\omega + \Delta\omega, S)$. Since the random process $\delta(S)$ is
considered as markovian, it can be argued~\citep{Lacey1993MNRAS.262..627L} that
$f^{(0)}(\omega_1, S_1\,|\,\omega_2, S_2) = 
f^{(0)}(\omega_1 - \omega_2, S_1 - S_2)$.
Expression for conditional PDF $f^{(0)}(\omega_2, S_2\,|\,\omega_1, S_1)$ can
be obtained via the formula of total probability
(\ref{eq:joint_prb_sink},\ref{eq:joint_prb_source}).

Let's modify the source and the sink of (\ref{eq:kinetic_ps_exact_usual}) in a
way to allow the mass ejection from the forming halos. Consider the probability
$f^{(0)}(\omega, S\,|\,\omega + \Delta\omega, S')\,dS$, which is a mass
fraction of fluctuations having variance range from $S$ to $S + dS$ and
collapsing by the  time $\omega$, after each absorbs  halos $S'$, which 
 existed before by the 
time $\omega + \Delta\omega$. As the sub-halos $S'$ may escape the
proto-halo, this fraction may decrease.

Let denote as $f(\omega,S)$ the unknown mass
function and to assume that the statistics of the fluctuations is not affected by the
sub-halos ejection occurring in other proto-halos (those which  collapsed
earlier). In this case, the  product
$[1 - \chi_\mathrm{ej}(\omega, S, S')]\,%
f^{(0)}(\omega, S\,|\,\omega + \Delta\omega, S')\,%
f(\omega + \Delta\omega,S')\,dS$
gives contribution to the mass of the halo $S$ by the  sub-halos $S'$,
which could not leave the parent. Then the mass fraction of all the halos that
form during interval $\Delta\omega$ is
\begin{multline}
  \label{eq:kinetic_source_basic}
  P_+(\omega + \Delta\omega, \omega, S)\,dS \equiv{}  \\
  {}\equiv dS \int_S^\infty dS'\,[1 - \chi_\mathrm{ej}(\omega, S, S')]\,
  f^{(0)}(\omega, S\,|\,\omega + \Delta\omega, S')\,
  f(\omega + \Delta\omega, S')  \;.
\end{multline}
Here $f(\omega + \Delta\omega, S')$ is the  mass function of the halos 
formed by the
time $\omega + \Delta\omega$. Expression (\ref{eq:kinetic_source_basic}) is the
first component of the source. It corresponds to the mass redistribution after
growth and virialization of the perturbations. Besides that, we need to take
into account those of the sub-halos which leave the proto-halos and return to
the population:
\begin{equation}
  P_{+,\mathrm{re}}(\omega + \Delta\omega, \omega, S) \equiv
  \int_0^S dS''\,\chi_\mathrm{ej}(\omega, S'', S)\,
  f^{(0)}(\omega, S''\,|\,\omega + \Delta\omega, S)\,
  f(\omega + \Delta\omega, S)  \;.
\end{equation}
This term must be incorporated into the source with negative sign.

The sink also has two components:
\begin{equation}
  P_-(\omega + \Delta\omega, \omega, S) \equiv
  f(\omega + \Delta\omega, S)\,
  Q_-(\omega + \Delta\omega, \omega, S)  \;,
\end{equation}
\begin{equation}
  P_{-,\mathrm{re}}(\omega + \Delta\omega, \omega, S) \equiv
  f(\omega + \Delta\omega, S)\,
  Q_{-,\mathrm{re}}(\omega + \Delta\omega, \omega, S)  \;,
\end{equation}
where
\begin{equation}
  Q_-(\omega + \Delta\omega, \omega, S) \equiv
  \int_0^S dS''\,[1 - \chi_\mathrm{ej}(\omega, S'', S)]\,
  f^{(0)}(\omega, S''\,|\,\omega + \Delta\omega, S)  \;,
\end{equation}
\begin{equation}
  Q_{-,\mathrm{re}}(\omega + \Delta\omega, \omega, S) \equiv
  \int_0^S dS''\,\chi_\mathrm{ej}(\omega, S'', S)\,
  f^{(0)}(\omega, S''\,|\,\omega + \Delta\omega, S)  \;.
\end{equation}

After all substitutions, the kinetic equation has the form:
\begin{equation}
  \label{eq:kinetic}
  - \DPfrac{f}{\omega} =
  \lim_{\Delta\omega \to 0+} \frac{1}{\Delta\omega}
      \left[ P_+ - P_{+,\mathrm{re}} - (P_- - P_{-,\mathrm{re}}) \right]  \;.
\end{equation}
It's easy to show by evaluation that
\begin{equation}
  \label{eq:normalization}
  \int_0^\infty dS\,(P_+ + P_{-,\mathrm{re}}) =
  \int_0^\infty dS\,(P_- + P_{+,\mathrm{re}}) = 1  \;.
\end{equation}
Expressions (\ref{eq:kinetic}) and (\ref{eq:normalization}) reveal that the
normalization of the PDF $f$ is conserved with the time. Note also that
$P_{+,\mathrm{re}} = P_{-,\mathrm{re}}$. Let's finally rewrite the equation
(\ref{eq:kinetic}) using Taylor expansion in $\Delta\omega$ and obtain then a recurrent procedure for calculation of the mass function:
\begin{equation}
  \label{eq:kinetic_calc}
  f(\omega, S) = f(\omega + \Delta\omega, S)
  + P_+(\omega + \Delta\omega, \omega, S)
  - f(\omega + \Delta\omega, S)\,Q_-(\omega + \Delta\omega, \omega, S)
  + \mathcal{O}(|\Delta\omega|^2)  \;.
\end{equation}

Neglecting the quadratic residue, the recurrent relation
(\ref{eq:kinetic_calc}) can be written as
\begin{equation}
  \label{eq:kinetic_integral}
  f(\omega, S) = \int_0^\infty dS'\,
  r(\omega, S\,|\,\omega + \Delta\omega, S')\,f(\omega + \Delta\omega, S')  \;,
\end{equation}
where
\begin{align}
  \label{eq:transition_probability}
  r(\omega, S\,|\,\omega + \Delta\omega, S') &=
  \theta(S' - S)\,q(\omega, S\,|\,\omega + \Delta\omega, S') +{}  \\
  &{}+ \delta_\mathrm{D}(S - S')
  - \delta_\mathrm{D}(S - S')
  \int_0^{S'} dS''\,q(\omega, S''\,|\,\omega + \Delta\omega, S')  \;,
\end{align}
and  
\begin{equation}
  q(\omega, S\,|\,\omega + \Delta\omega, S') \equiv
  [1 - \chi_\mathrm{ej}(\omega, S, S')]\,
  f^{(0)}(\omega, S\,|\,\omega + \Delta\omega, S')  \;.
\end{equation}
The representation (\ref{eq:kinetic_integral}) is useful for numerical solution
of the kinetic equation by Monte-Carlo method. One step of the solution procedure
may look like this:
\begin{itemize}
  \item[(a)] generate the random value $S'$ with  initial PDF $f(\omega, S')$;
  \item[(b)] generate the random value $S$ according to the transition PDF
    (\ref{eq:transition_probability});
  \item[(c)] redefine $\omega - \Delta\omega \mapsto \omega$,
    $S \mapsto S'$ and go to (b).
\end{itemize}
A cumulative transfer distribution function is more convenient for this algorithm
than the PDF:
\begin{equation}
  \int_0^S dS''\,r(\omega, S''\,|\,\omega + \Delta\omega, S') =
  \left\{
  \begin{aligned}
    & \int_0^S dS''\,q(\omega, S''\,|\,\omega + \Delta\omega, S')
    \;,\quad S \leqslant S'  \\
    & 1 \;,\quad S > S'
  \end{aligned}
  \right.
\end{equation}

\begin{figure}[!ht]
  \centering
  \includegraphics{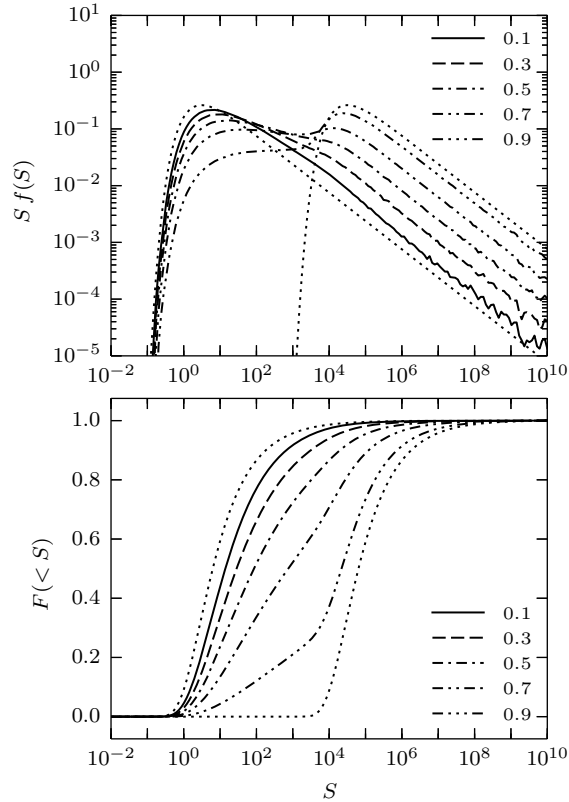}
  \caption{The mass function (PDF and the cumulative) for the case $\chi_\mathrm{ej} \equiv \operatorname{const}$. The rightmost dashed line designates the initial distribution at $z = 100$ for all runs. The leftmost dashed line designates the final distribution $z = 0$) for the case $\chi_\mathrm{ej} = 0$. The intermediate curves correspond to various $\chi_\mathrm{ej}$ (see the legend on the plot).}
  \label{fig:excursion_const}
\end{figure}

Examples of how the sub-halo ejection affects the halo mass function are shown
in Fig.~\ref{fig:excursion_const}.
In these calculations, the escape probability was assumed constant. Also the EPS mass function was used as the
initial PDF. It is clearly seen that for larger
escape probability $\chi_\mathrm{ej}$ the \glqq{tail}\grqq\ of
the low mass end (high $S$ values) of the mass function becomes  heavier , i.e. the evolution of
the population effectively slows down as $\chi_\mathrm{ej}$ increases.

\subsection{Choice of initial conditions}

In the kinetic approach presented above it is possible to use an appropriate arbitrary
initial mass function. Thus, the problem of the choice of the initial
conditions arises. According to the modern cosmological theory,
matter-dominated era started at $z \approx 3 \times 10^3$
\citep{Gorbunov2006vvedenie_hot_bing_bang}. According to the formal relation
(\ref{eq:variance_interval}), structures with the variances
$S \sim 10^6 \dots 10^7$ or masses $\log(m/M_\odot) \ll -100$  formed
at that era (see Appendix). It is obvious, however, that the masses of the
structures can not be lower than the mass of a hypothetical dark matter
particle. If we take the value $m_\mathrm{DM} = 100$~GeV
$\sim 10^{-55} M_\odot$ as the mass of the dark matter particle (this value
corresponds to $S_\mathrm{DM} \approx 10^4$), then the earliest structures may
form at the redshifts as low as $z \sim 100$ (see Eq.
(\ref{eq:variance_interval})), and it is unlikely for structures to form at the
higher redshifts. Finally, recalling the sub-halo escape model, we see that
the sub-halos can not escape at the redshifts greater than $250$ (see
Fig.~\ref{fig:ejfrac_lowm}). Considering all these restrictions, we assume the
redshift value $z_\mathrm{i} \equiv 300$ as the initial moment for all
calculations below. As the initial mass function we assume the cumulative
distribution function
\begin{equation}
  \label{eq:initial_mass_function}
  F(\omega_\mathrm{i}, <S) =
  \left\{
  \begin{aligned}
    & F^{(0)}(\omega_\mathrm{i}, <S) \;,\quad S \leqslant S_\mathrm{DM}  \\
    & 1 \;,\quad S > S_\mathrm{DM}  \;,
  \end{aligned}
  \right.
\end{equation}
where $\omega_\mathrm{i} = \delta_\mathrm{c}/D(z_\mathrm{i})$. Note that
$F^{(0)}(\omega_\mathrm{i}, <S_\mathrm{DM}) \sim 10^{-7}$.

\begin{figure}[!ht]
  \centering
  \includegraphics{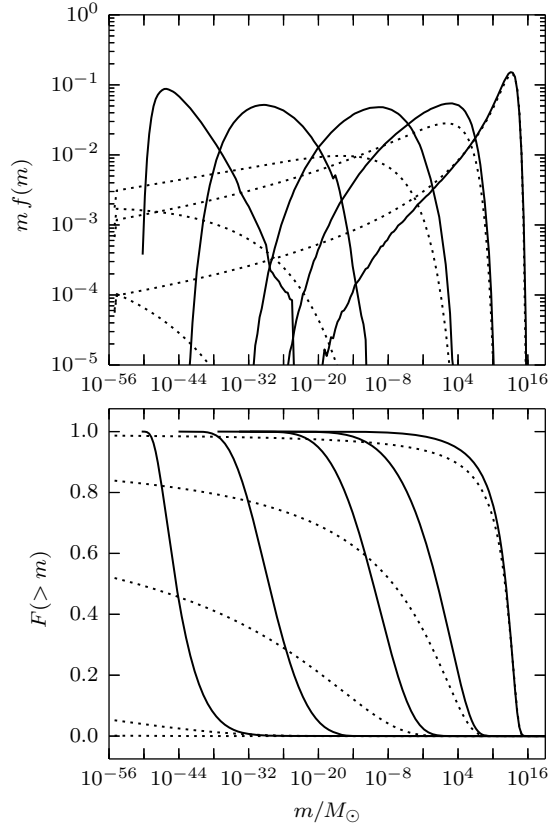}
  \caption{Mass function (PDF and the cumulative) for the case $\chi_\mathrm{ej} = 0$ and the initial conditions Eq.~(\ref{eq:initial_mass_function}). The dotted lines designates the PS mass functions. Five epochs are represented in the plot (from left to right on both panels): $z = 250$, $z = 150$, $z = 50$, $z = 15$, and  $z = 0$.}
  \label{fig:excursion_dps-ps}
\end{figure}

In Fig.~\ref{fig:excursion_dps-ps}
the mass functions in the model without sub-halo ejection ($\chi_\mathrm{ej} = 0$) are plotted for the different redshifts. The initial mass
function (\ref{eq:initial_mass_function}) peaks at the greater masses
relatively to the peak of the EPS mass function (at $z = 300$ the EPS peaks at
$S \sim 10^5$), so the mass function at the subsequent times is also shifted to
the higher masses. However, at $z \lesssim 15$ the massive end of the
distribution coincides with the EPS model. In general, the low mass end of the
mass function is suppressed at all redshifts, because of absence of the
structures of mass $m_\mathrm{DM}$ and lower.

Obvious consequence of using the initial mass function suggested in this section
is that the formation of the structures with masses lower than $m_\mathrm{DM}$
is impossible. This property may be used for the analysis of the resolution effects
of cosmological codes built on top of the N-body or grid methods. The lowest
possible mass scale in such applications should correspond to the spatial
resolution of the numerical method.

\subsection{Account for the background structure}

Effect of the low density background structure (the supercluster or the void) can
be considered in our model. Such a structure may be set in terms of the
statistical constraints for the field of the overdensity
fluctuations~\citep{Kurbatov2014ARep...58..386K}. Applying the constraints
leads to considerable change in statistics of modes of the fluctuations. E.g. the statistics is not
more spatially uniform, the perturbations field \glqq{feels}\grqq\ the size and
shape of the background structure. Thus, the halo mass function also
changes. In the EPS theory the effects of the background was considered first
by \citet{Bond1991ApJ...379..440B,Bower1991MNRAS.248..332B}, and
investigated further by \citet{Mo1996MNRAS.282..347M,Sheth1999MNRAS.308..119S} and others. When
deviation of density of the background structure from uniform distribution is low, and the structure is
large enough in all directions, the effect of the background presence can be
accounted by a   simple approach. Let's denote
$S_\mathrm{B} \equiv S(m_\mathrm{B})$ the variance of the overdensity
fluctuations averaged over the background structure's mass scale
$m_\mathrm{B}$, then the variance of the field over the mass scale
$m \ll m_\mathrm{B}$ placed deep inside the background structure is approximately
$S(m) - S_\mathrm{B}$
\citep{Hoffman1991ApJ...380L...5H,Kurbatov2014ARep...58..386K}. The same considerations
are true for the velocity fluctuations field also (this fact was already
utilized when the escape criterion (\ref{eq:relative_velocity_variance}) was
obtained).

\citet{Bond1991ApJ...379..440B,Bower1991MNRAS.248..332B} have shown
that the large scale background structure can be easily applied to the PS
mass function Eq.~(\ref{eq:eps_mass_function}) by formal substitutions of the kind
\begin{equation}
  \label{eq:background_shift}
  \omega \mapsto \omega - \omega_\mathrm{B}
  \;,\qquad
  S \mapsto S - S_\mathrm{B},
\end{equation}
where $\omega_\mathrm{B}$ is the background overdensity calculated via the
linear law (\ref{eq:linear_overdensity}), and reduced to the unit growth factor:
\begin{equation}
  \omega_\mathrm{B} =
  \frac{D(z = 0)}{D(z_\mathrm{i})}\,\Delta(z_\mathrm{i}) =
  \frac{\delta_\mathrm{c}}{D(z_\mathrm{f})}  \;.
\end{equation}
The second equality in this expression allows  to define an overdense structure
which collapses by the moment $z_\mathrm{f}$. The equivalent description can be
reached in the kinetic approach proposed in the previous section, if the substitutions
(\ref{eq:background_shift}) are made in the transfer PDF
$f^{(0)}(\omega, S\,|\,\omega + \Delta\omega, S')$.

The same substitutions may be utilized in the model with kinematic effects. It
should be noted however, that the substitutions (\ref{eq:background_shift})
must not be applied to escape probability $\chi_\mathrm{ej}(\omega, S, S')$,
because in fact the probability depends on the redshift and the halo mass, but
not the overdensity threshold and the overdensity variance. Moreover, the
velocity variance already enters  Eq.~(\ref{eq:escape_fraction}), as the
difference, hence its value, does not change as the velocity field varies, if
these changes have the form (\ref{eq:background_shift}). Finally, the transfer
cross section is
\begin{equation}
  q(\omega, S\,|\,\omega + \Delta\omega, S') =
  [1 - \chi_\mathrm{ej}(\omega, S, S')]\,
  f^{(0)}(\omega - \omega_\mathrm{B}, S - S_\mathrm{B}\,|\,
  \omega - \omega_\mathrm{B} + \Delta\omega, S' - S_\mathrm{B})  \;,
\end{equation}
where $\omega_\mathrm{B} \equiv \delta_\mathrm{B}/D$. Note that the initial
conditions (\ref{eq:initial_mass_function}) must not be transformed by
(\ref{eq:background_shift}).

\section{Virialized matter in the local Universe}

Mass counts of the galaxies, virialized groups of galaxies, and clusters, made
by many authors reveal the deficiency of the mass inside virialized objects, up to the
factor three (see references in \citet{Makarov2011MNRAS.412.2498M}) compared to
predictions of $\Lambda$CDM model. According to the catalog \citet{Makarov2011MNRAS.412.2498M}, local density parameter
of the matter
\begin{equation}
  \label{eq:local_matter_density}
  \Omega_\mathrm{m} \equiv \frac{3 M_\mathrm{m,tot}}%
        {4\pi\rho_\mathrm{cr,0} D^3}  \;,
\end{equation}
may be estimated as $0.08 \pm 0.02$. Here $M_\mathrm{m,tot}$ is the total mass
of the matter inside a sphere of radius $D$. The authors of the catalog
proposed several explanations of the deficiency. The most plausible of them is the
suggestion about possibility that the essential part of the dark matter in
the Universe (about $2/3$) is scattered outside the virial or collapsing
regions, being distributed diffusively or concentrated in dark
\glqq{clumps}\grqq~\citep{Makarov2011MNRAS.412.2498M}. Certain evidence
for existence of the dark clumps may be provided by
the observations of weak lensing events
\citep{Natarajan2004ApJ...617L..13N,Jee2005ApJ...618...46J}, and the properties of
the disturbed dwarf galaxies \citep{Karachentsev2006A&A...451..817K}.

The estimate $\Omega_\mathrm{m} = 0.08 \pm 0.02$ was obtained by
\citet{Makarov2011MNRAS.412.2498M} using the total mass of all virialized
groups of galaxies having velocities up to $3500$~km$/$s. To estimate the
distances and masses the authors used Hubble parameter's value
$73$~km$/$s$\,$Mpc. Thus the distance $D$ can be estimated as
$3500/73 \approx 48$~Mpc, and the estimate of the total mass of the matter
(given $\Omega_\mathrm{m,0} = 0.28$) is $1.92 \times 10^{16} M_\odot$. The
value of the Hubble parameter adopted in the present paper is
$H_0 = 67.3$~km$/$s$\,$Mpc, whence $D = 52$~Mpc, and the total mass is thus
$M_\mathrm{m,tot} \approx 2.33 \times 10^{16} M_\odot$. The variation of the
Hubble parameter causes just a minor correction to the estimated galactic
masses, about $8\%$. The estimate of the local matter density parameter
changes more significantly and gives (when the masses of galaxies are corrected
also)
$\Omega_\mathrm{m} = 0.068 \pm 0.017 = (0.22 \pm 0.05) \Omega_\mathrm{m,0}$.
The claims of  \citet{Makarov2011MNRAS.412.2498M} were
based on observational data of the virialized groups having masses above
$\approx 3 \times 10^{10} M_\odot$, and containing at least two
galaxies. Completeness of the sample was $82\%$. The mass distribution function
of the virialized groups (see Fig.~8 in \citet{Makarov2011MNRAS.412.2498M}) showed
decrease  of the number of objects with masses lower than $10^{13} M_\odot$. This can be
caused by the absence of the field galaxies in the sample. In the further analysis we
will use just a fraction of the sample composed by the objects as massive as
$10^{13} M_\odot$ and above. The reason is that this fraction should more
closely match the mass distribution of the dark matter gravitational
condensations.

\begin{figure}[!ht]
  \centering
  \includegraphics{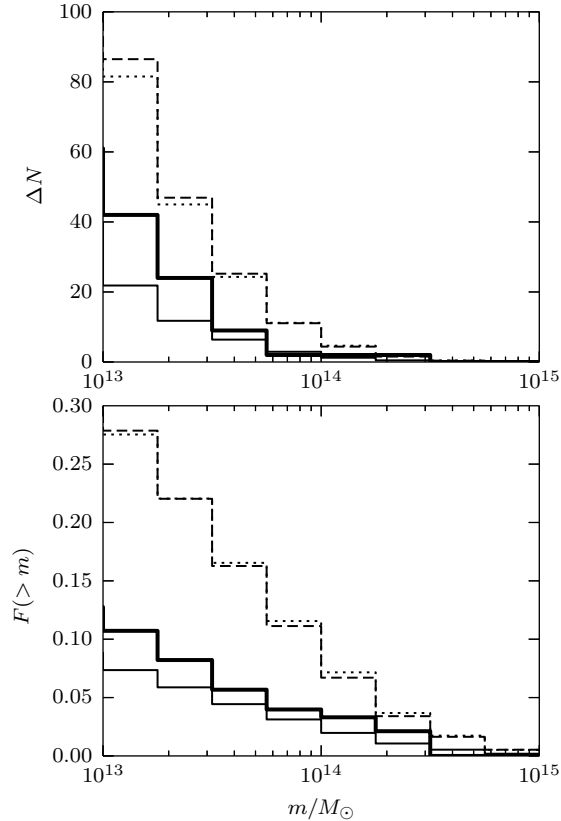}
  \caption{Mass function of the virialized structures with masses $> 10^{13} M_\odot$:  t\citet{Makarov2011MNRAS.412.2498M} (thick solid line), the model with sub-halo ejection (thin solid line),  the model without ejection, but with the initial conditions 
Eq.~(\ref{eq:initial_mass_function}) (dashed line), the PS model (dotted line).}
  \label{fig:mk}
\end{figure}

According to EPS distribution, the total mass fraction of the dark matter halos
more massive than $10^{13} M_\odot$ is
$M_\mathrm{m}(>10^{13} M_\odot) / M_\mathrm{m,tot} \approx 0.27$, i.e. more
than $2/3$ of all the dark matter settles in the halos with masses below 
$10^{13} M_\odot$ (see Fig.~\ref{fig:mk}).
Calculation of the model presented in Section~3.2, performed neglecting the
sub-halo escape but using the initial conditions
(\ref{eq:initial_mass_function}) leads to approximately the same distribution
as the EPS predicts (dashed line on Fig.~\ref{fig:mk}). The observational data
(thick solid line on Fig.~\ref{fig:mk}) show the deficit of the virialized
halos, up to the factor two and a half comparing to the EPS prediction:
$M_\mathrm{m}(>10^{13} M_\odot) / M_\mathrm{m,tot} \approx 0.11$.

\begin{figure}[!ht]
  \centering
  \includegraphics{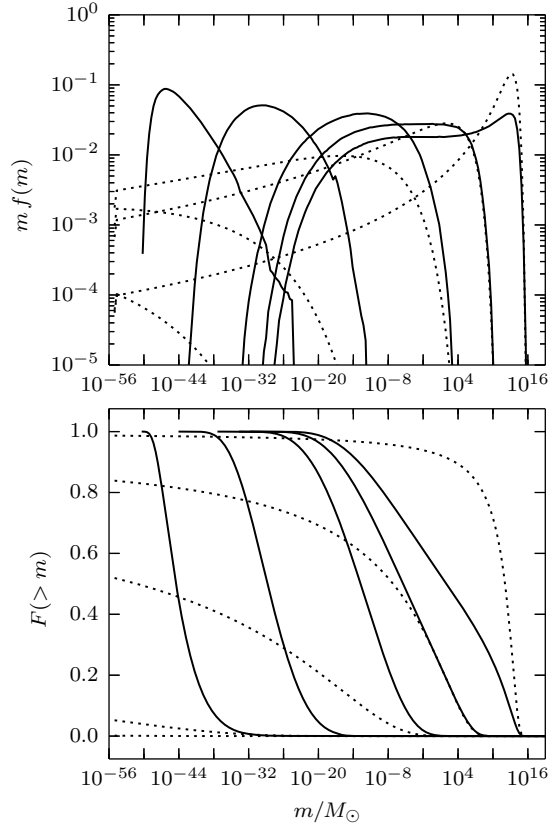}
  \caption{Mass function (PDF and the cumulative) calculated with Eq.~(\ref{eq:escape_fraction}). Lines indicate the same as in  Fig.~\ref{fig:excursion_dps-ps}.}
  \label{fig:excursion_dps-ej}
\end{figure}

Accounting for the effect of sub-halos escape in the presented model, gives the cumulative fraction of all the massive halos about $0.08$, i.e. slightly lower than observed. The most part of the matter is not incorporated into the structures as massive as $10^{13} M_\odot$ and above but becomes distributed nearly uniformly in 
structures with  masses greater than $10^{-10} M_\odot$ (see Fig.~\ref{fig:excursion_dps-ej}). 
Quite a good agreement between the observed and the theoretical distribution functions of the massive halos is mainly due to the fact that the differential distribution functions for both cases nearly coincide at the masses from $3 \times 10^{13} M_\odot$ and above (see the upper plot in Fig.~\ref{fig:mk}), whereas two left bins on this plot show twice as large deficit of the halos predicted the theoretical model suggested in this study.
While the uncertainties in the estimates of the groups' masses can be as high as $20\%$ \citep{Makarov2011MNRAS.412.2498M}, the 
average values of the observed mass function are systematically higher than those of the theoretical one. This may be explained in two ways:
\begin{itemize}
  \item[(i)] Average matter density in the vicinities of the local Supercluster seems to be higher than cosmological value~\citep{Makarov2011MNRAS.412.2498M}. As a consequence of this, the massive tail of the halo mass PDF should be heavier
    \citep{Mo1996MNRAS.282..347M,Kurbatov2014ARep...58..386K}.
  \item[(ii)] The proposed effect of the mass decrease of proto-halos may be overestimated as we did not account for, e.g., accretion of the matter on the proto-halo from its environment. \citet{Diemand2007ApJ...667..859D} have shown that halo may accrete up to $20\%$ of mass after virialization has ended, i,e., accretion may partially compensate the escape process leading to effective decrease of the escape probability $\chi_\mathrm{ej}$.
\end{itemize}

In general, a good agreement between the observed mass function   \citep{Makarov2011MNRAS.412.2498M} and predictions of the model proposed in this paper
should be noted.

\section{Conclusions}

In the present paper the influence of the dark matter random velocities on formation of the dark halos population is considered. It is shown that this kinematic effect leads to 
the escape of the significant fraction of sub-halos (up to $30\%$).
However, at the redshifts greater than $250$ the escape is negligible. At the moderate and small redshifts the escape probability is higher for sub-halos of lower masses.

We suggested a model of the dark halo population formation, based on the EPS theory. The model was built in the frame of kinetic approach where the source and the sink  explicitly describe the hierarchical merging process and the kinematic effects, and potentially may be used to account environmental effects. Special initial conditions were proposed to use with the kinetic equation where the lowest possible halo mass exists corresponding to the dark halo particle's mass. The consequences of these initial conditions are systematic shift of the low-mass end of the mass distribution function towards high masses. However, this shift is almost invisible in the high-mass end at lower redshifts.  The consequences of the sub-halo escape is the redistribution of the massive halos toward lower masses. If no escape is included, the EPS mass function is the invariant solution of the kinetic equation.

The model developed in this paper allowed to explain quantitatively the observable deficit
 of the virialized objects in the Local Universe, at least for groups of galaxies with masses greater than $10^{13} M_\odot$. The missing matter is distributed over the low-mass halos down to the lowest limit.

We need to note that the kinetic approach proposed in this paper, with the initial conditions limiting the lowest possible halo mass, may be used for analysis of the resolution effects in the cosmological numerical codes.

\section{Appendix A}

Let us obtain an approximate expression for the overdensity variance reduced to unity scale factor:
\begin{equation}
  S(M) = \int d^3k\,|\tilde{W}(\vecb{k}, M)|^2\,P(k)  \;.
\end{equation}
Assume \glqq{k-sharp}\grqq\ filter with the Fourier image
$\tilde{W}(\vecb{k}, M) = \theta(1 - k X)$, where
$X = ( 3 M/4\pi \Omega_\mathrm{m,0} \rho_\mathrm{cr,0} )^{1/3}$. Then the variance is
\begin{equation}
  S(M) = \frac{1}{2\pi^2} \int_0^{X_M^{-1}} dk\,k^2 P(k)  \;.
\end{equation}
The overdensity power spectrum for large wavenumbers may be written as~\citep{Bardeen1986ApJ...304...15B}
\begin{equation}
  P \approx A k^{n_\mathrm{S}}\,\frac{\ln^2(2.34 k/\Gamma)}{k^4}  \;,
\end{equation}
where constant $A$ is defined by normalization $\sigma_8$; $\Gamma = h^2 \Omega_\mathrm{m,0}$~Mpc$^{-1}$. Assume $M \ll M_\ast$, then
\begin{equation}
  S(M) \approx S(M_\ast)
  + \frac{A}{2\pi^2 (1-n_\mathrm{S})^3 \Gamma^{1-n_\mathrm{S}}}
  \left[ \frac{ (1-n_\mathrm{S})^2 \ln^2\varkappa
      + 2 (1-n_\mathrm{S}) \ln\varkappa + 2}%
    {\varkappa^{1-n_\mathrm{S}}}
    \right]_{(\Gamma X)^{-1}}^{(\Gamma X_\ast)^{-1}}  \;.
\end{equation}
Making all substitutions, we obtain
\begin{equation}
  \label{eq:overdensity_variance_approx}
  S(M) \approx 1.29
  + 10^4
  \left[ 2.052
  - \left( 0.001 \lg^2 M_{14} - 0.062 \lg M_{14} + 2 \right) M_{14}^{0.01323}
  \right]  \;,
\end{equation}
where $M_{14} \equiv M / 10^{14} M_\odot$.

The variance of the relative velocity of a low mass sub-halo inside the parent halo can be estimated in a similar way. The variance on a mass scale $M$ reduced to the unit growth factor is
\begin{equation}
  S_v(M) =
  \int d^3k\,|\tilde{W}(\vecb{k}, M)|^2\,\frac{P(k)}{k^2}
  = S_v(0)
  - \frac{1}{2\pi^2} \int_{X^{-1}}^\infty dk\,P(k)  \;.
\end{equation}
The relative velocity variance for low-mass sub-halos is then
\begin{equation}
  S_v(0) - S_v(M) \approx
  \frac{A}{2\pi^2 \Gamma^{3-n_\mathrm{S}}}
  \left. \frac{(3-n_\mathrm{S})^2 \ln^2\varkappa
      + 2 (3-n_\mathrm{S}) \ln\varkappa + 2}%
    {(3-n_\mathrm{S})^3 \varkappa^{3-n_\mathrm{S}}}
    \right|_{\varkappa = (\Gamma X)^{-1}}  \;.
\end{equation}
Assuming that the power spectrum has unit spectral index $n_\mathrm{S} = 1$, after all substitutions we have:
\begin{equation}
  \label{eq:relative_velocity_stddev_approx}
  \left[ S_v(0) - S_v(M) \right]^{1/2} \approx
  1.75 \left[ 1.178 \log^2 M_{14} - 3.55 \log M_{14} + 3.17 \right]^{1/2}
  M_{14}^{1/3}  \;,
\end{equation}
The expression in the square root term is a slowly changing monotonic decreasing function of the proto-halo mass. Approximately, it is $14$ for $M = 10^2 M_\odot$ and unity for $M = 10^{14} M_\odot$.


\bibliography{paper}
\bibliographystyle{plainnat}

\end{document}